# A Machine-Learned Interatomic Potential Free-Energy Surface for the First Step of MIL-101(Cr) Secondary Building Unit Formation

*Orlando A. Mendible-Barreto and Yamil J. Colón**

Department of Chemical and Biomolecular Engineering, University of Notre Dame, Notre Dame, IN, 46556

## Abstract

Metal-organic framework (MOF) self-assembly mechanisms remain poorly characterized because *ab initio* molecular dynamics (AIMD) are accurate but too computationally expensive to converge the free-energy surfaces (FES) that govern secondary building unit (SBU) nucleation and growth. This work addresses that limitation for the first step of MIL-101(Cr) SBU formation. Using MACE-POLAR-M, a long-range-aware equivariant machine-learned interatomic potential fine-tuned on a compact DFT reference dataset built from exploratory metadynamics, this work obtains a converged 2D FES for this step at a level of theory (ωB97M-V/def2-TZVPP) at which AIMD would be prohibitively expensive. The pre-trained model samples relevant configurations but gets the thermodynamics wrong, predicting an endothermic reaction and a false global minimum. Fine-tuning corrects both, recovering the expected exothermic process and the correct product basin in agreement with proposed literature mechanisms. The resulting FES also refines the mechanism. It shows that water release from the chromium center proceeds stepwise, through two sequential energy barriers, rather than the single concerted step originally proposed. We find that model accuracy is specific to the reaction coordinate and does not extend to the dissociated fragment configurations excluded from fine-tuning, a direct consequence of the training-data choice that protects accuracy along the reaction path for the FES prediction task. Together, these results establish a validated, computationally tractable route to obtain free-energy surfaces for MOF formation chemistry, transferable to other reactions where AIMD remains prohibitive, and identify training-data composition as the key determinant of which reaction properties a fine-tuned checkpoint can reliably describe.

# 1 Introduction

Metal–organic frameworks (MOFs) are porous crystalline materials assembled from metal ions or clusters connected by organic linkers. The large number of combinations of these building blocks gives MOFs their modular chemistry, allowing pore size, surface chemistry, and framework topology to be tuned without heavily compromising one property for another, compared to other materials.[1] This tunability has made MOFs useful across gas storage[2,3] and separation[4,5], heterogeneous catalysis[6,7], drug delivery[8,9], and other applications[10–12], with more than 100,000 distinct MOF structures reported to be successfully synthesized to date[13]. MOF synthesis remains highly uncertain despite this practical relevance, as evidenced by the difference between theoretically viable MOFs ($>10^6$)[14,15] versus experimentally synthesizable ones[14,15]. The lack of understanding of the fundamental science underlying MOF formation mechanisms means that realizing thermodynamically stable structures reproducibly with specific topology or morphology remains a challenge[16,17] and synthesis conditions are still largely identified by empirical trial and error rather than by a predictive mechanistic model.[18,19] A thorough and precise knowledge of MOF synthesis pathways is currently inaccessible because of the complexity of metal–anion–solvent interactions[20], which involve many competing non-covalent and covalent interactions on similar energy scales[21], and because of the high number of competing, simultaneous events that occur during MOF formation.[22–24] Classical and *ab initio* simulation methods are both limited in their ability to capture the rare events that drive MOF formation: classical force fields lack the electronic-structure detail needed to describe bond-forming and bond-breaking interactions accurately, and *ab initio* molecular dynamics (AIMD) simulations, while accurate, are too computationally expensive to sample the rare events and long timescales required to describe all of these interactions for a full formation pathway.[25–27]

MOF nucleation and the formation of secondary building units (SBUs) are rate-limiting steps that remain poorly characterized at a mechanistic level.[17] Chromium terephthalate MIL-101 is a widely studied MOF since it combines a large unit-cell volume (approximately 702,000 $Å^3$)[28], extra-large pore diameters (29 to 34 Å)[28], and a specific surface area greater than 4,000 $m^2/g$ [29], properties that have made it a benchmark framework for gas storage and as a host for embedding secondary nanomaterials.[30] The formation of MIL-101's chromium-oxo SBU serves as a mechanistic model for how metal and linker precursors assemble into a crystalline framework.[31,32] Cantu et al.[33] provided the first computational, stepwise mechanism for a MIL-101 SBU, using CP2K[34] with the PBE exchange-correlation functional to run AIMD and exploratory metadynamics at 300 K, which identified metastable configurations along the reaction path.[33] That study identified the terephthalate linker's carboxylate groups joining chromium(III) centers prior to formation of the bridging oxo group as a key mechanistic step, and reported a potential energy barrier of approximately 35 kcal/mol, associated with a high-to-low spin transition as a third chromium-linker unit joins the growing cluster. Cantu et al. explicitly did not attempt to reconstruct a converged free-energy surface (FES) for any step or reaction of the mechanism due to the computational cost of doing so with AIMD-driven metadynamics.[33]

Machine-learned interatomic potentials (MLIPs) offer a path to closing this gap between classical and *ab initio* molecular simulations by providing the accuracy of *ab initio* methods at a fraction of the computational cost.[35–37] Equivariant message-passing architectures such as MACE reached high accuracy at a computational cost orders of magnitude below electronic-structure methods by encoding the relevant physical symmetries directly into the model,[38,39] and large-scale MACE foundation models trained across broad regions of chemical space, such as MACE-MP-0[40], made this accuracy available without system-specific training. MACE-POLAR[41] extended this architecture with explicit long-range electrostatics and a polarizable induction scheme, allowing a single foundation model to be applied to systems with appreciable charge transfer and polarization, such as the metal-linker-solvent clusters involved in SBU formation. To date, however, this model has not been tested on FES calculation tasks via enhanced sampling simulations for SBU systems. Enabling the use of MLIPs to study the self-assembly of MOFs will close a gap that has limited understanding of the mechanistic driving forces that promote the formation of these materials.[24] Realizing this potential for a specific reaction required that the pre-trained foundation model be reliably specialized ("fine-tuned") without losing its general-purpose accuracy elsewhere in configurational space.[42]

A recent systematic study of MACE fine-tuning strategies compared three approaches. Naïve (full-parameter) fine-tuning gave the best convergence for narrow, single-system applications. Low-rank adaptation (LoRA)[43] reached comparable accuracy while updating a small fraction of the model's parameters, but drifted furthest from the pre-trained distribution. Replay-based multi-head fine-tuning was the only one of the three that consistently preserved out-of-distribution robustness.[44] That study benchmarked these trade-offs on lithium electrolytes, ice polymorphs, aqueous NaCl, $S_N2$ reactions, and biomolecules in the SPICE dataset[45], but not for a MOF-formation reaction, where the reaction coordinate involves large, correlated changes in metal-center charge and spin state on top of the known concerns about catastrophic forgetting related to MLIP fine-tuning.[44] Prior work applying MLIPs to MOFs has so far focused on phase behavior and simulation benchmarking rather than on the bond-forming chemistry of formation itself. Vandenhaute et al.[46] combined an equivariant NequIP potential with parallelized metadynamics and on-the-fly DFT labeling to construct transferable, thermodynamically accurate MLIPs for UiO-66(Zr) and the flexible MIL-53(Al) framework from only a few hundred quantum-mechanical evaluations per material, and used the resulting potentials to resolve a breathing phase transition inaccessible to direct AIMD.[46] More recently, a systematic benchmark of universal MLIPs across geometry optimization, molecular dynamics stability, and host-guest interactions in MOFs found substantial variation in reliability across models and tasks, underscoring that accuracy on one MOF simulation task does not guarantee accuracy on another.[46] For a chemically different framework family, reactive and machine-learned potentials have been used to study self-assembly directly: partially reactive force field metadynamics identified the synthesis stage at which zeolitic imidazolate framework (ZIF) polymorph selection occurs, finding that pre-nucleation clusters were already polymorph-dependent[47], and a MACE-architecture MLIP was subsequently used alongside that reactive force field to generate configurations for phase classification in the same ZIF system[48], while a separate high-throughput crystal-structure-prediction study used custom-trained MLIPs to sample millions of candidate ZIF packing arrangements[49].

This work replicates the mechanisms presented by Cantu et al. describing the early-stage self-assembly of the MIL-101(Cr) SBU, using an MLIP fine-tuned at a higher level of theory than what was previously used.[33] The chemical equation for this reaction is presented below:

$$[Cr(H_2O)_6]^{3+} + BDC^{-} \rightarrow [Cr(H_2O)_4\text{–}BDC]^{2+} + 2\ H_2O \qquad (1)$$

This is the first reaction in the mechanism, starting with a Cr(+3) ion in an octahedral configuration with six water molecules coordinated within the metal's first solvation shell and a partially deprotonated BDC linker (see state a in Figure 1). Two water molecules left the solvation shell of the metal atom to form an intermediate state (state b in Figure 1), and the oxygens of the BDC linker took their place to form the product of the reaction (state c in Figure 1). This reaction serves as a controlled test case for demonstrating the usability of fine-tuned MLIPs and enhanced sampling methods to discover fundamental knowledge that will help better understand MOF self-assembly mechanisms. The reactant, intermediate, and product proposed by Cantu et al. are shown in Figure 1 along with the calculated potential energy surface (PES) obtained from nudged elastic band (NEB) calculations. Cantu et al. calculated and reported the binding energy for this reaction as a value of −84.5 kcal/mol.

Our primary goal is to reproduce the mechanisms described for this reaction and calculate the free energy barriers that separate metastable configurations using a well-sampled FES determined using a MLIP trained at a higher level of theory. The success of this goal will enable a calculation that would be prohibitively expensive with *ab initio* methods and provide a means to discover fundamental knowledge about the mechanisms that govern the formation of this material. Additionally, we probe the ability of the fine-tuned models to calculate the binding energy of this reaction and analyze the effect of the used training data on accuracy of the specific tasks the models are used for.

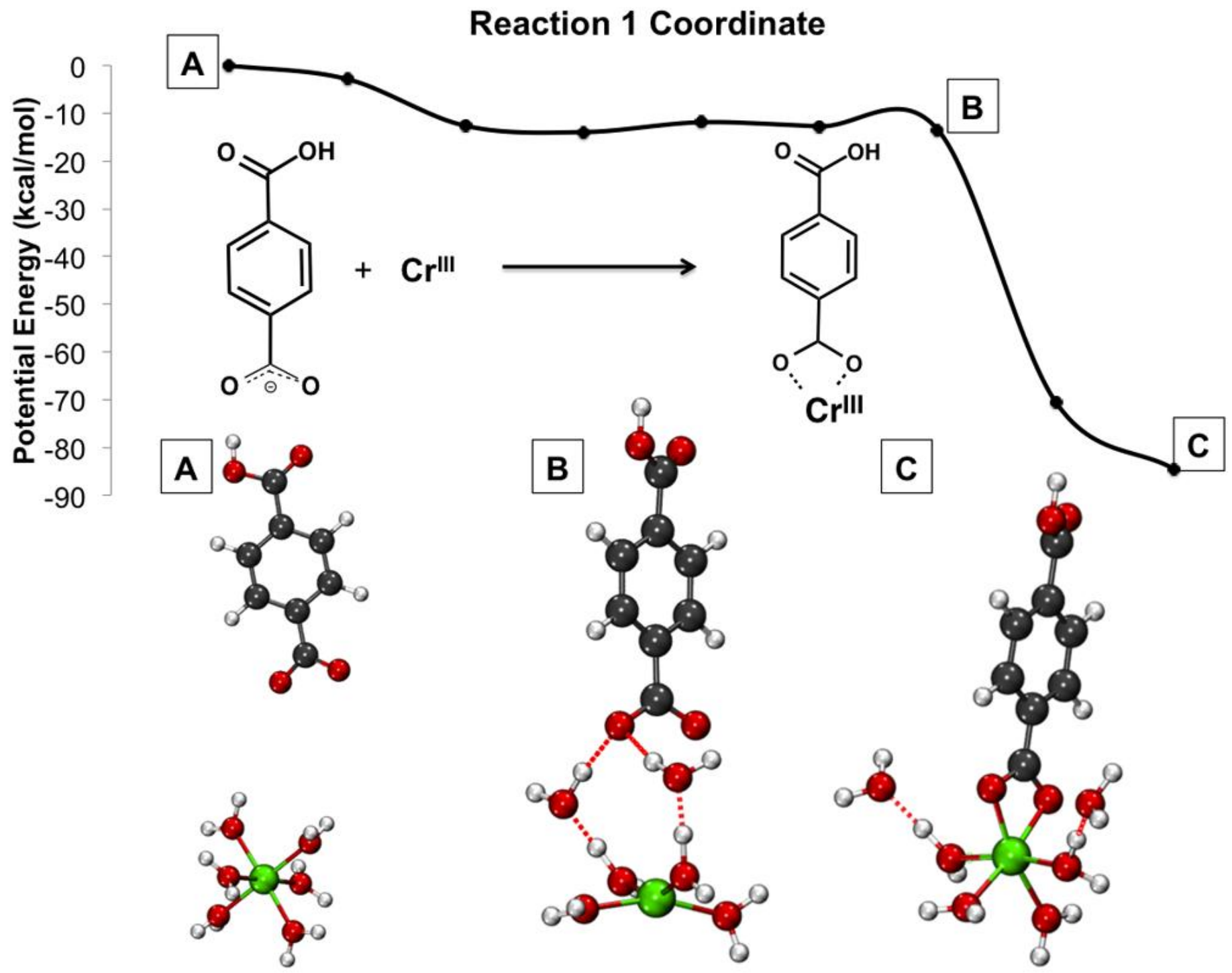

*Figure 1. Reaction 1 for the formation of MIL-101 SBUs presented by Cantu et al.[33] Green is Cr(III), red is O, gray is C, white is H. Reproduced from Ref. [33] with permission from the American Chemical Society.*

# 2 Methodology

## 2.1 Computational pipeline overview

The reaction was treated with a six-stage pipeline, executed with the pre-trained MACE-POLAR-M[41,50] foundation model and then repeated with fine-tuned variants of the same model:

**1. Reactant preparation.** Initial geometry was prepared using Packmol[51] via DynaMate2[52] to generate individual configurations of the metal cluster and the linker, and then combine them to obtain the reactant state identified as A in Figure 1. The pretrained model was used to relax the configuration, which was then used in the following steps.

**2. Unbiased stability testing.** A 3 ns unbiased NVT molecular dynamics simulation (additional simulation details in Section 2.3) at 300 K was run from the relaxed structure to confirm that the pre-trained model preserved the expected coordination environment and did not exhibit unphysical configurations or energy drifts.

**3. Exploratory metadynamics.** Well-tempered metadynamics (additional simulation details in Section 2.4) with physically motivated collective variables (coordination numbers between the Cr atom and water molecules and between the Cr atom and BDC oxygens) was used to drive the system along the reaction coordinate and generate a diverse set of configurations spanning the reactant, intermediates, and product basins.

**4. Reference data generation and model fine-tuning.** Local free-energy minima ("basins") were identified on the exploratory metadynamics surface, representative configurations were extracted, and single-point DFT reference calculations (Section 2.5) were computed at the same level of theory used to train the foundation model. This reference dataset was used to fine-tune MACE-POLAR-M by three distinct strategies (naïve fine-tuning, low-rank adaptation (LoRA), and multi-head replay fine-tuning, Section 2.6). Each fine-tuned model was evaluated against the pre-trained model and the DFT reference. Two types of datasets were evaluated, one containing the fragments (i.e, individual molecules in vacuum) and one without fragments (nofrag), only containing the configurations obtained from the exploratory metadynamics. All datasets included 10 images of the individual atom types. Additional details about the configurations used in each dataset are available in section 1 of the SI.

**5. Free-energy surface (FES) calculation.** Production well-tempered metadynamics runs were performed independently with the pre-trained model and with fine-tuned models to obtain a two-dimensional FES along the sampled collective variables. Production runs ran for 7 ns, and convergence analysis plots are available in the SI.

**6. Binding energy calculation.** The binding energy of the reaction was first reproduced at the same level of theory as Cantu et al. to ensure that similar configurations were used in this test. Then, this value was calculated using the pre-trained and fine-tuned models (Section 2.7)

## 2.2 Simulation software and environment

All molecular dynamics were performed with the Atomistic Simulation Environment[53] (ASE) as the simulation driver, using the MACE-POLAR-M[41,50] machine-learned interatomic potential (a long-range-aware MACE[38] variant with explicit atomic multipole/polarization terms) as the energy and force calculator with CUDA-accelerated equivariant operations[54,55] and GPU execution optimization. Enhanced sampling was applied through the PLUMED[56] plugin, coupled to ASE. Density functional theory reference calculations for fine-tuning were performed with ORCA[57] (Section 2.5). All workflows were executed within a single dedicated conda environment to ensure consistent package versions (ASE, MACE, PLUMED, CUDA) across all simulation stages. The conda environment and necessary files are available in this project's repository.

## 2.3 ASE simulation parameters

Geometry optimizations of reactant and product endpoints used the ASE LBFGS optimizer with the pre-trained MACE-POLAR-M force field using double precision (float64), a convergence threshold on the maximum atomic force of 0.05 eV/Å, for up to 400 optimization steps. All molecular dynamics were propagated with ASE's Langevin[58] thermostat at 300 K, a timestep of 0.5 fs, and a friction coefficient of 0.1 $fs^{-1}$, using the MACE-POLAR-M calculator in single precision (float32). This temperature matches the value used in Cantu et al.'s exploratory AIMD metadynamics. All molecular dynamics and metadynamics simulations were conducted in a periodic cubic box with side lengths of 30.5 Å. The spin (S) and charge of the system are 1.5 and 2, respectively, given that Cantu et al. (their SI Table S3) report that both the $[Cr(H_2O)_6]^{3+}$ reactant and the $[Cr(H_2O)_4–BDC]^{2+}$ product independently favor a quartet (S = 1.5) ground state over the doublet, by approximately 26 kcal/mol for each species. Stability testing simulations ran for 3 ns (unbiased) and 7 ns (exploratory metadynamics), while production metadynamics simulations ran for 7 ns.

## 2.4 PLUMED collective variables and enhanced-sampling parameters

Reaction progress was monitored and biased using two primary collective variables (CVs): coordination numbers between the Cr atom and water molecules (coord_Os) and between the Cr atom and BDC oxygens (coord_bdc). These CVs were calculated using the switching function presented in equation 2, where n and m are exponential parameters that determine the steepness of the inner and outer decay, respectively.[56] $r_{ij}$ is the distance between atoms *i* and *j*, $r_0$ is the reference distance scaling parameter, and $d_0$ is a shift parameter, which we set to 0.

$$s_{ij} = \frac{1 - \left(\frac{r_{ij} - d_0}{r_0}\right)^n}{1 - \left(\frac{r_{ij} - d_0}{r_0}\right)^m} \tag{2}$$

The switching function parameters used for these CVs are r=2.90, n=10, and m=18. Additional details about the analysis for the selection of these parameters is provided in the SI. Well-tempered metadynamics simulations were used to bias the reaction coordinate using a Gaussian height of 0.3 kcal/mol, a bias factor of 9, and sigma of 0.15 and 0.05 for coord_Os and coord_bdc. In addition, harmonic upper/lower walls were used to maintain a distance between 3.0 and 12 Å to ensure the reactants remained close enough to react. The files used to run these simulations are included in this project's repository, and additional details are provided in the SI.

## 2.5 ORCA reference calculations

Single-point density functional theory reference energies and gradients were computed with ORCA[57] (version 6.1.0) at the ωB97M-V/def2-TZVPP[59] level of theory, using the resolution-of-identity chain-of-spheres exchange approximation (RIJCOSX)[60] with the def2/J auxiliary basis[61], tight SCF convergence, and a dense integration grid (DEFGRID3)[62]. This is the same level of theory used to train the MACE-POLAR-M foundation model, which kept the reference data consistent with the pre-trained model.

Reference configurations were selected directly from the exploratory metadynamics trajectory. The metadynamics bias potential was reconstructed on a discretized free-energy grid, seven local free-energy minima (“basins”) were identified computationally, and the ten representative trajectory frames closest to each basin minimum in CV space were selected for single-point calculation (70 configurations total). Isolated-atom and reference-fragment single-point calculations (at the same level of theory) were additionally computed to support atomic-energy referencing and binding-energy validation, and the resulting labeled dataset (DFT energies and forces) was partitioned into training and validation sets. An independent, non-overlapping set of held-out configurations was assembled (ten further frames per basin) by the same procedure to serve as a test set for generalization checks (Section 3). Full basin-selection statistics, dataset composition, and split fractions are given in SI Table S1.

## 2.6 MACE fine-tuning approaches

The MACE-POLAR-M foundation model was fine-tuned on the DFT reference dataset using three distinct strategies, all initiated from the same pre-trained checkpoint via the mace_run_train training utility:

Naïve *fine-tuning:* unrestricted fine-tuning of all foundation-model weights on the reaction-specific reference dataset.

*LoRA (low-rank adaptation) fine-tuning:* the foundation-model weights were frozen, and small low-rank update matrices ( SI Table S2) were trained in their place, reducing the set of parameters that received gradient updates while retaining the pre-trained representation.

*Multi-head (replay) fine-tuning:* the reaction-specific reference data was fine-tuned alongside a replay subset of the original foundation-model training data (sampled without additional filtering), using a multi-head output architecture intended to mitigate catastrophic forgetting of the general-purpose pre-trained potential.[42]

All three strategies shared the same equivariant CUDA acceleration (enable_cueq) and single precision training; energy/force loss weighting was not uniform across strategies (Naïve used reduced weights relative to LoRA and Multi-head), and strategy-specific hyperparameters (learning rate, epoch count, early-stopping patience, EMA decay, LoRA rank/alpha, and replay-set size) are given in SI Table S2.

## 2.7 Binding energy calculation

As an independent validation, the Reaction 1 binding energy reported by Cantu et al.[33] was reproduced with periodic DFT at their level of theory (PBE functional, CP2K34), and the pre-trained and fine-tuned MACE-POLAR-M models were tested for whether they could independently recover this reference geometry and energetics. Full methodology and results are given in SI Section S5.

# 3 Results and Discussion

## 3.1 Stability of the pre-trained model

We first tested the pre-trained foundation model for dynamical stability before generating any reference data for the early-stage self-assembly of the MIL-101(Cr) SBU, which included the chromium hexahydrate node and the BDC linker in vacuum. Unbiased and metadynamics simulations were run, whose energy, temperature, and sampled CVs are summarized in Figure 2. They consist of a 3 ns unbiased NVT run and a 7 ns exploratory well-tempered metadynamics run biasing the Cr–$O_{water}$ and Cr–$O_{BDC}$ coordination numbers (Section 2.4). Panels a and b show that both simulations equilibrated to a stable plateau of approximately 330 K, modestly above the nominal 300 K thermostat set point, with no measurable total-energy drift over the full trajectory and no unphysical bond dissociation or ligand loss. Histograms in panels c and d demonstrate that during the unbiased simulation, the Cr atom remained coordinated to six water molecules and that no waters left the solvation shell of the metal ion and were replaced with BDC's oxygens. They also indicate that the metadynamics simulations sample configurations of interest for the reaction, where two water molecules left the Cr solvation shell and were substituted by the carboxylic group of the BDC linker.

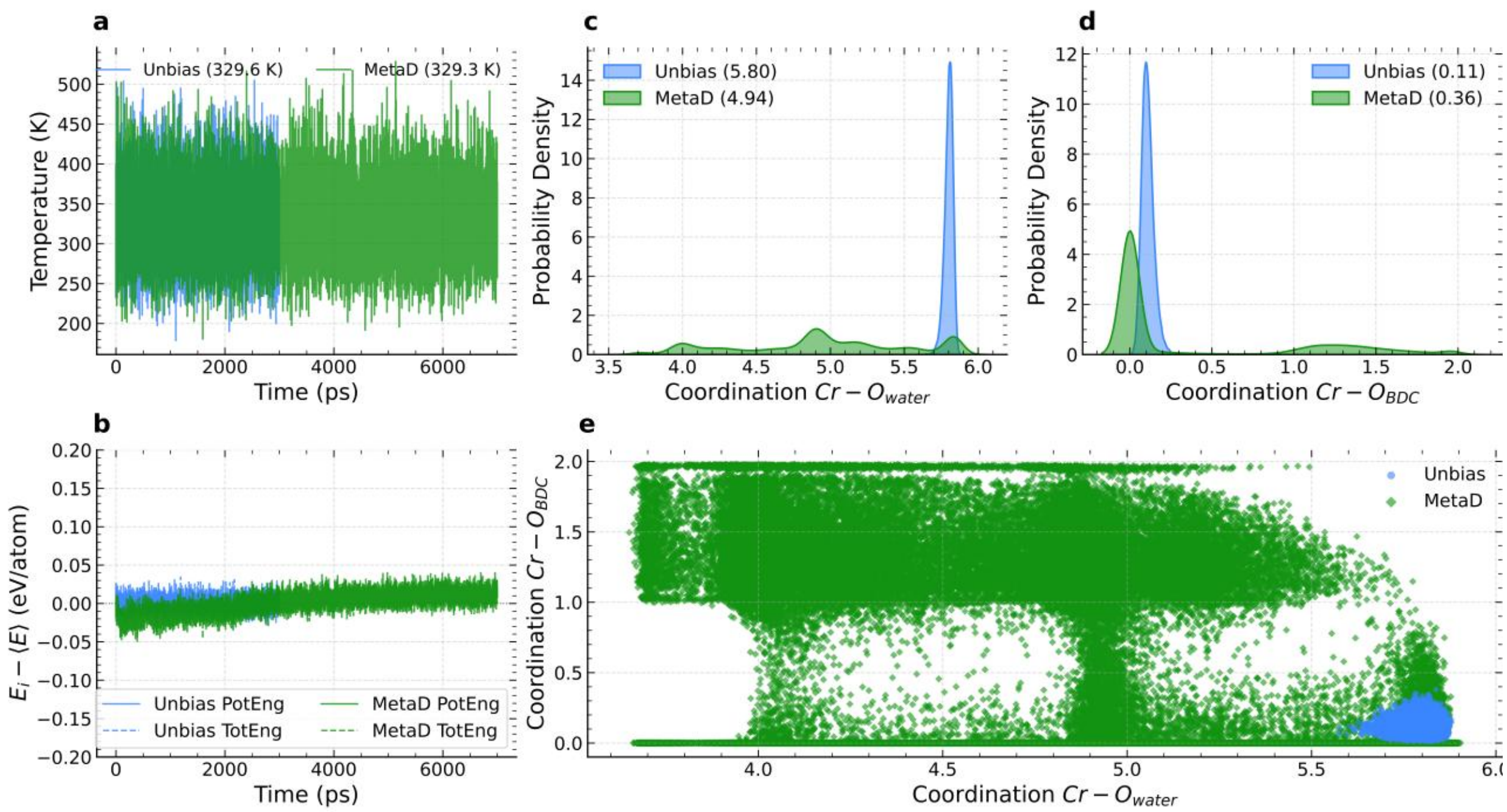

*Figure 2. Stability of the pre-trained MACE-POLAR-M model on Reaction 1. (a) Temperature vs. time for unbiased NVT and exploratory metadynamics runs. (b) Normalized potential and total energy vs. time for the same runs. (c)*

*Probability distribution of the Cr–O_water coordination number, sampled during both the unbiased and exploratory metadynamics runs. (d) Probability distribution of the Cr–O_BDC coordination number, sampled during both runs. (e) Scatter plot of the two CVs, showing the explored region of configuration space.*

This interpretation is further validated in panel e, where each point represents the combination of CVs for sampled configurations, and shows that sampling of the unbiased only sampled the region where the Cr ion is solvated by six water molecules and was stuck in this basin, while the exploratory metadynamics run visited a broad and chemically sensible region of the Cr–$O_{water}$/Cr–$O_{BDC}$ coordination plane, spanning the reactant, intermediate, and product states. These trajectories formed the basis for the basin/frame selection used to build the DFT reference dataset (Section 2.5).

The temperature offset is systematic: every simulation in this study, run with the same float32 MACE-POLAR-M foundation model and cuEquivariance-accelerated kernels, converges to 330 K rather than the 300 K set point, and none show progressive drift once equilibrated. This behavior is consistent with a growing body of work on precision effects of MLIPs in molecular dynamics.[63–65] Reduced-precision arithmetic and GPU-optimized equivariant kernels, including the cuEquivariance backend used here, can introduce small numerical inconsistencies between a model's predicted energy and the forces it returns. Because the forces are then not the exact analytic gradient of the energy under finite-precision execution, the dynamics absorb a small amount of injected numerical noise each step, which a Langevin thermostat continuously removes; the net effect is a stable, elevated apparent temperature rather than a trajectory that fails to conserve energy or drifts without bound. Explicit precision benchmarks of MACE with the cuEquivariance backend in NVT and NPT water simulations report exactly this kind of numerically induced shift, with thermodynamic observables remaining reproducible run-to-run but offset from full double-precision references.[63] Float32 precision limits on energy conservation, and their direct effect on the simulated temperature of a neural-network-potential trajectory, have been characterized quantitatively in other MLIP implementations[64], and the broader consequences of small deviations from strictly conservative forces for molecular dynamics stability are an active area of methodological research.[65] Because the offset is constant across every run and model compared in this work rather than a progressive drift, it does not compromise the qualitative comparisons between the pre-trained and fine-tuned models that are the focus of this study, and is treated here as a known precision limitation of the current MLIP simulation setup.

Panel b reports each run's potential and total energy as a deviation from its own trajectory mean, $E_i - \langle E \rangle$, rather than on a shared absolute scale. Read this way, the unbiased run's energy is essentially flat, consistent with its confinement to a single coordination basin (Figure 2 panels c–e). The metadynamics run's larger excursions in panel b are not the same precision-related effect discussed above: they track the bias-driven exploration of distinct Cr-coordination environments, each with a different characteristic potential energy. The potential energy is strongly correlated with the Cr–$O_{BDC}$ coordination number, falling by approximately 0.02 eV/atom as the system moves from the water-solvated state (coord_bdc ≈ 0) toward BDC-bound configurations (coord_bdc ≈ 2). The unbiased and metadynamics runs also differ in duration by design: the unbiased run needs only to confirm dynamical stability within a single basin, which panels c–e already show it does not leave, while the exploratory metadynamics run requires substantially more simulated time to adequately sample the broader reactant-intermediate-product region of CV space used to build the DFT reference dataset (Section 2.5).

The sampled configurations from the metadynamics run were used to obtain a proxy for the FES via the Boltzmann distribution, thereby identifying configurations that represent regions of high sampling (i.e., where the free energy is minimized) and mapping the configurational space of the reaction of interest. Figure 3 shows the snapshots of the main configurations of interest obtained from the metadynamics simulation with the pretrained MACE-POLAR model. These configurations were used as inputs to single-point calculations with ORCA and subsequently as training data for fine-tuning the models discussed below. Additional metastable configurations, obtained from this simulation and included in the fine-tuning and test datasets but not revisited during the fine-tuned models' production metadynamics runs, are excluded from the five-state (R, EB1, Int, EB2, P) reaction-pathway analysis in Section 3.3 but are presented in Figure S1 and discussed separately in Section 2 of the SI.

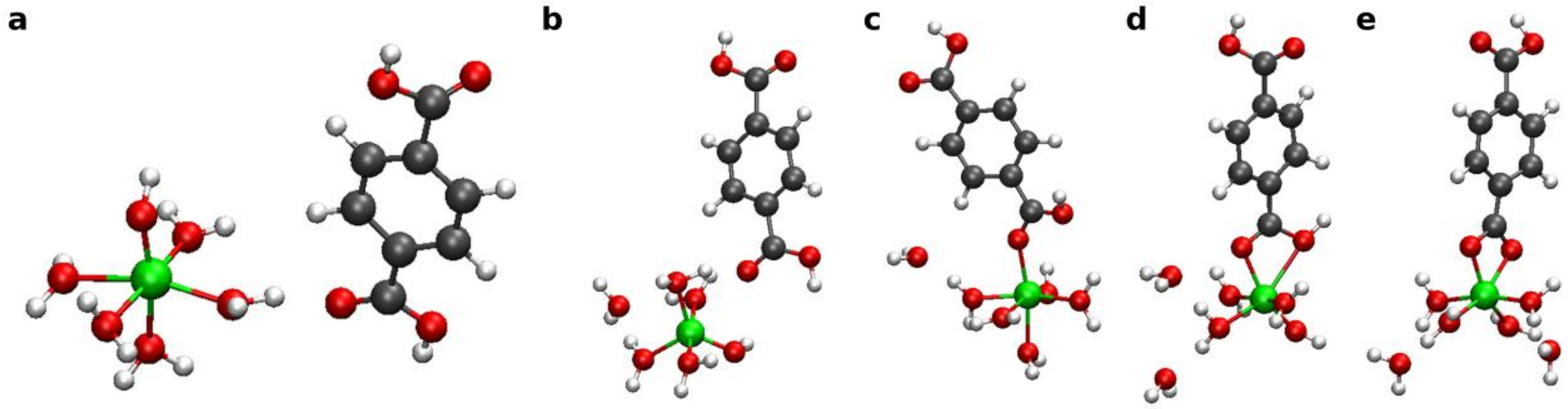

*Figure 3. Configurations representing the mechanism described as Reaction 1 by Cantu et al. for the self-assembly of MIL-101 SBUs, obtained from the metadynamics simulation with the pretrained MACE-POLAR model. Panels a through e correspond to the reactant (R), energy barrier 1 (EB1), intermediate (Int), energy barrier 2 (EB2), and product (P), respectively. Green is Cr, red is O, grey is C, and white is H.*

## 3.2 Validation of the pre-trained and fine-tuned models

Obtaining a trustworthy production FES first required determining which fine-tuning strategy yielded a sufficiently accurate model for deployment on metadynamics simulations. How well each strategy reproduced the DFT reference energies and forces on the reaction coordinate was tested for the training and test datasets.

For the training dataset, the pre-trained foundation model, evaluated zero-shot, disagreed with the DFT reference at the ωB97M-V/def2-TZVPP level of theory by an energy mean absolute error (MAE) of 19.4 meV/atom (Pearson r = 0.89) and a force MAE of 108.1 meV/Å ($R^2$ = 0.93), consistent with the model never having seen similar environments during its initial training. This result, however, highlights the usability and out-of-the-box performance of the MACE-POLAR model to describe interactions of a charged and complex system involving a charge transfer and polarization effects. All three fine-tuning strategies closed most of this accuracy gap as shown by the evaluation on the training dataset: Naïve fine tuning reached 1.2 meV/atom (r = 0.998) energy accuracy and 48.3 meV/Å ($R^2$ = 0.975) force accuracy, and multi-head was comparable on energies (2.4 meV/atom, r = 0.997) and best on forces (41.0 meV/Å, $R^2$ = 0.970). LoRA improved on the pre-trained baseline (9.1 meV/atom, r = 0.967; 61.3 meV/Å, $R^2$ = 0.972) but did not match the two full-weight strategies. The same ranking is held on a test dataset of 70 held-out configurations unseen during fine-tuning. The results of the evaluations on the test data are presented as energy and forces parity plots in Figure 4.

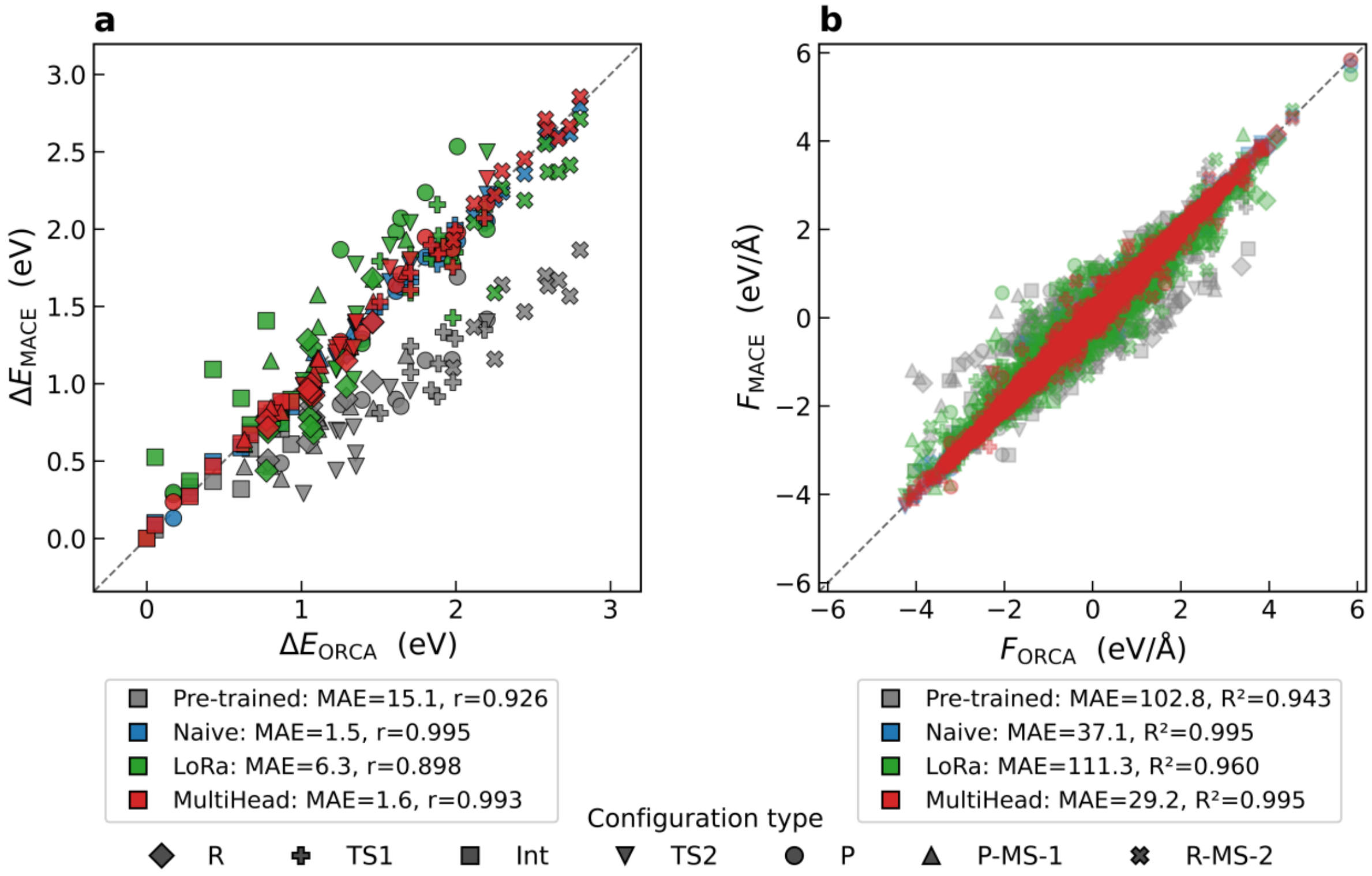

*Figure 4. MACE-POLAR predicted vs. DFT-reference (ωB97M-V/def2-TZVPP) energy (a) and atomic forces (b) for the pre-trained model and each fine-tuned strategy, evaluated on the 70-structure held-out test set (Section 2.6). Mean Absolute Error (MAE) is reported in meV/atom and meV/Å for the energy and force, respectively.*

The MAE for the energy and forces of each model were: pre-trained 15.1 meV/atom / 102.8 meV/Å, Naïve fine-tuned 1.5 meV/atom / 37.1 meV/Å, multi-head fine-tuned 1.6 meV/atom / 29.2 meV/Å, and LoRA fine-tuned 6.3 meV/atom / 111.3 meV/Å. For the Naïve and MultiHead fine-tuned models, the force accuracy improved on the test set relative to the training-adjacent numbers, indicating these models generalized along the reaction pathway rather than memorizing the sampled frames. LoRA is the exception: its test-set force MAE (111.3 meV/Å) is 82% higher than its own training-adjacent value (61.3 meV/Å) and worse than the pre-trained baseline's test-set force MAE (102.8 meV/Å). This result suggests a genuine generalization failure specific to this strategy, not observed by the other two models. As a result, the LoRa-fine-tuned model was not used for the metadynamics production runs.

Parity plots comparing DFT-reference and MACE-POLAR predicted energies and forces for models fine-tuned on datasets that included dissociated fragment configurations ($[Cr(H2O)6]^{3+}$, BDC-, and isolated atoms, each carrying a different net charge and spin than the reaction coordinate of interest) showed a decrease accuracy for the configurations related to the reaction coordinate, and additionally the results showed that the size of the effect is strategy-dependent: on-reaction-coordinate force MAE increases by 120% for Naïve (12.9 to 28.4 meV/Å) but only 2% for MultiHead (11.3 to 11.5 meV/Å) when fragment configurations are included in training, while on-reaction-coordinate energy MAE degrades meaningfully for both strategies (Naïve: 44%, 0.85 to 1.22 meV/atom; MultiHead: 85%, 0.77 to 1.42 meV/atom). Because the pre-trained foundation model is already parametrized to handle a broad range of charge and spin states, fine-tuning it on a small, reaction-specific dataset spanning multiple additional charge/spin states (the dissociated $[Cr(H_2O)_6]^{3+}$ fragment, the BDC- fragment, the $Cr(H_2O)_4–BDC]^{2+}$ complex and isolated atoms) pulls the weights away from that general-purpose parametrization, degrading accuracy on the primary reaction coordinate task through catastrophic forgetting.[42] For this reason, the metadynamics production runs reported below use only the fine-tuned models without the fragments, whose training data was restricted to configurations sharing the same overall charge and spin as the reaction coordinate itself.

## 3.3 Free-energy surface comparisons between Pre-trained and Fine-tuned MLIPs

Fine-tuning reshaped the reaction thermodynamics: the pre-trained model predicted an endothermic reaction (product 15.80 kcal/mol above reactant), while both fine-tuned models predicted an exothermic reaction (product 13.46-16.29 kcal/mol below reactant). This sign flip in the overall thermodynamics, rather than a magnitude correction, and the reshaping of the FES are the most important results for this paper's fine-tuning argument. The free-energy landscape reshaped visibly between the pre-trained (Figure 5a) and fine-tuned production runs (Figure 5b and 5c). Beyond this sign flip, both fine-tuned models resolved different basin depths and positions along the Cr–water/Cr–BDC coordination plane relative to the pre-trained surface (Figures 5d and 5e).

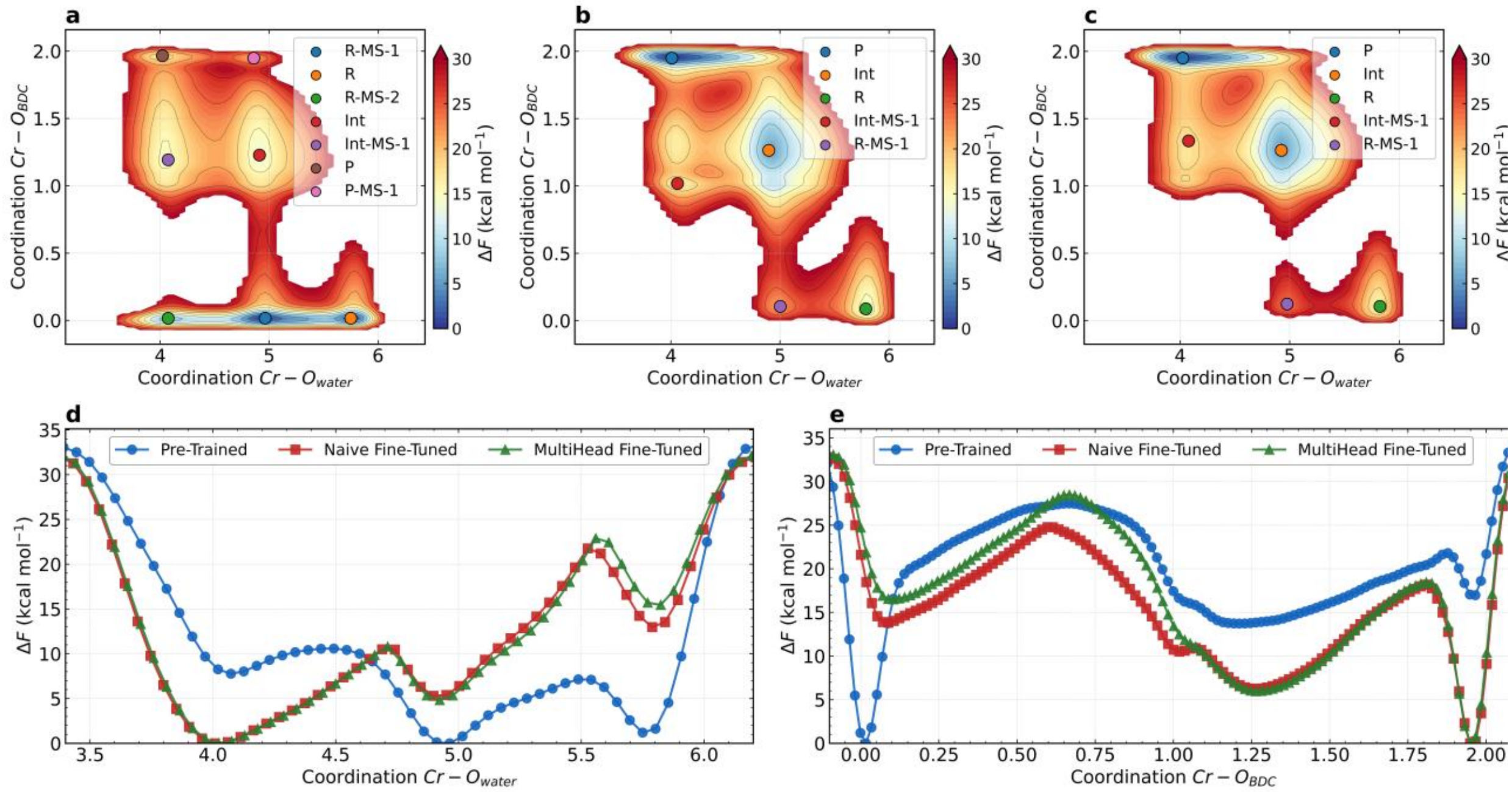

*Figure 5. 2D free-energy surfaces (x-axis: Cr–$O_{water}$ coordination number; y-axis: Cr–$O_{BDC}$ coordination number) from independent production well-tempered metadynamics runs with (a) the pre-trained model, (b) the Naïve fine-tuned model, and (c) the multi-head fine-tuned model. 1D projections along Cr–$O_{water}$ and Cr–$O_{BDC}$ CN collective variables in (d) and (e), respectively.*

The pretrained model identifies the global minimum as a configuration in which the Cr atom is coordinated to five water molecules and coordinated to no BDC linker, with a local minimum for configurations with six water molecules coordinated, and an energy barrier of ~7 kcal/mol to move between these two basins (These configurations are available in Section 2 of the SI). This six-water configuration is what this work refers to as "R" (reactant) throughout the paper: the literal hexaaqua structure, defined by the same collective-variable target for all three models, rather than each model's predicted global free-energy minimum. For the pre-trained model, R therefore sits 1.2 kcal/mol above its own (chemically incorrect) global minimum, a metastable state discussed in the SI. The Cr(+3) atom is known to prefer an octahedral arrangement with water molecules, so this shows that the pre-trained model is not accurate enough to estimate thermodynamic barriers, but it can sample the correct configurations. Interestingly, the pre-trained model found a local minimum corresponding to a water coordination of four when no BDC is coordinated. Fine-tuned models did not observe this configuration, indicating that atomic forces were properly updated to avoid exploring it, a result that aligns with the results reported by Cantu et al.[33] Moreover, the FES calculated from both fine-tuned models agree with the mechanisms reported by Cantu et al., where the global minimum corresponds to the product state, and where the rate-limiting step corresponds to the release of water molecules from the solvation shell of the metal atom when the linker is approaching it. In contrast, however, to the mechanisms presented in their work, where the two waters are first released to open the space for the BDC, the free energy surface and trajectories of this work indicate that a single water molecule is first released to open the space for a single BDC oxygen, generating the intermediate configuration. Then, the second water molecule is released, opening space for the linker to form a complete bidentate configuration with the metal ion. Convergence analysis plots are available in the SI.

Figure 6 summarizes the free-energy barriers and basins obtained by each model to facilitate the analysis of the results.

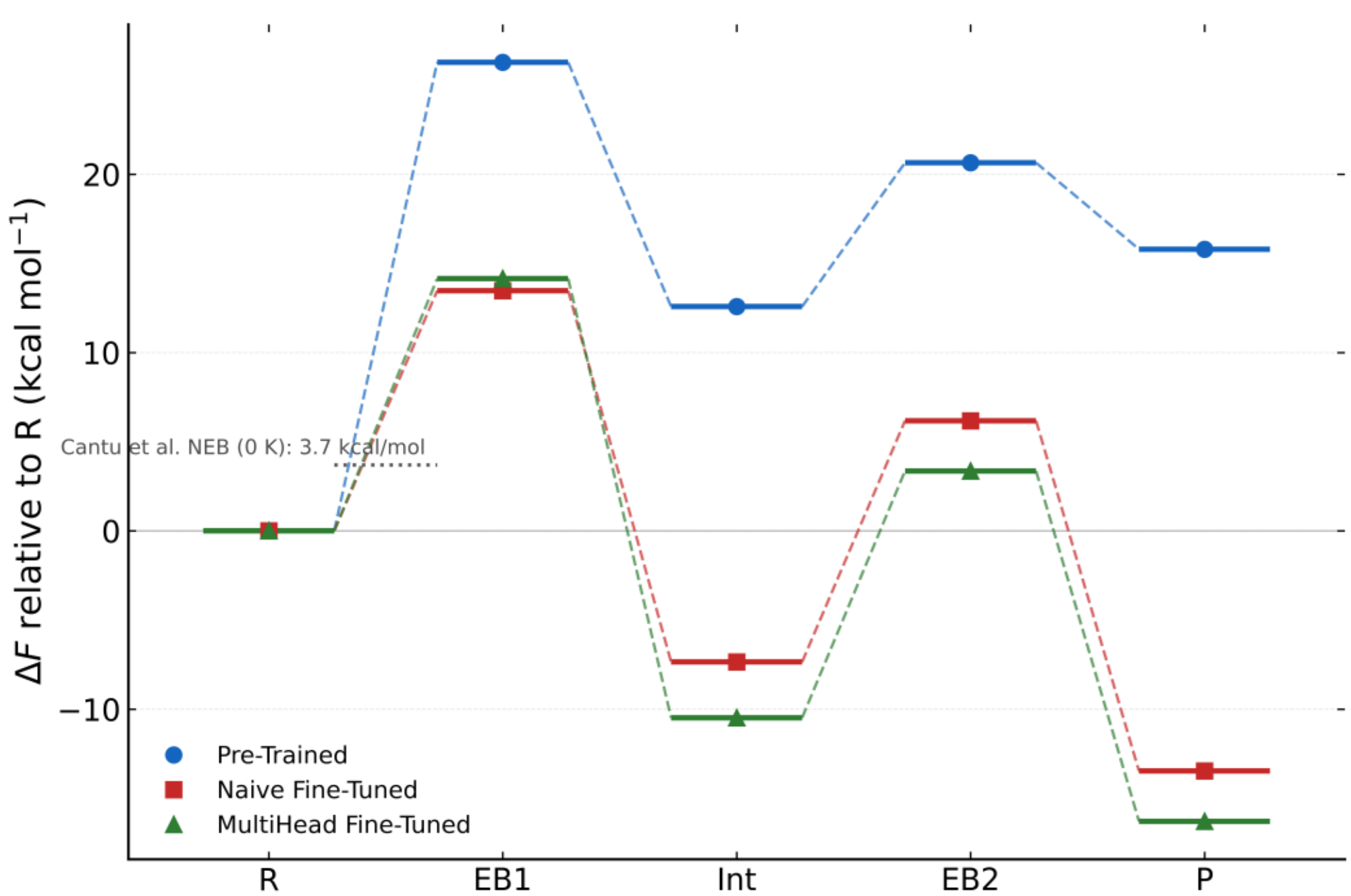

*Figure 6. Free-energy profile: basins and barriers by model. Free energy (ΔF, kcal mol⁻¹) of the reactant (R), first energy barrier (EB1), intermediate (Int), second energy barrier (EB2), and product (P) basins, relative to R (the hexaaqua reactant, defined identically for all three models). The dotted gray line indicates the energy barrier reported by Cantu et al. Solid segments mark each basin's free energy; dashed lines connect adjacent basins as a visual guide only, not a computed minimum-energy path.*

The global free-energy minima identified by the pretrained model did not match those identified by the fine-tuned models. The pretrained model identified a chromium configuration with five coordinated water molecules and no coordinated BDC as the global minimum, whereas the fine-tuned models identified the global minimum as the configuration in which four water molecules coordinate the chromium ion alongside a bidentate BDC linker, corresponding to the preferred Cr-octahedral product configuration. The free-energy surfaces predicted by the fine-tuned models agreed with the mechanism proposed by Cantu et al.: the highest barrier of the reaction corresponded to a water molecule leaving the coordination shell of the Cr atom (EB1), which all three models identified as the highest barrier, though the pretrained model underpredicted the barrier from the intermediate state to EB2 relative to the fine-tuned models (8.07 vs. 13.53-13.84 kcal/mol), which instead predicted barriers of more similar magnitude for EB1 and EB2. Although the pretrained model explored the configurational space and visited configurations that are part of the physically accessible landscape, its energies and forces did not describe the FES accurately and identified an incorrect global minimum.

Cantu et al. also reported a barrier for this specific step: their SI Table S.1 lists an energy barrier of 3.7 kcal/mol for Reaction 1, confirmed via climbing image nudge elastic band (CI-NEB) in the quartet spin state (SI Table S.4), distinct from the ~35 kcal/mol barrier discussed in the Introduction, which corresponds to a later step in the SBU-assembly mechanism (a high-to-low spin transition as a third chromium-linker unit joins the growing cluster), not to Reaction 1 itself. The fine-tuned models' R→EB1 free-energy barriers (13.5-14.2 kcal/mol) are roughly 3.6-3.8x higher than this reported value, and the pre-trained model's (26.3 kcal/mol) higher still, at roughly 7.1x. Four factors likely contribute to this gap. First, Cantu et al.'s barrier was obtained with the PBE exchange-correlation functional. Benchmarked against high-level coupled-cluster references for transition-metal reaction barriers specifically (the MOBH35 database[66] of 35 organometallic reactions), semi-local GGA functionals such as PBE are not among the best-performing methods, whereas ωB97M-V, the functional underlying the DFT reference data used to fine-tune MACE-POLAR-M in this work, is one of the most accurate functionals tested against this benchmark (mean absolute deviation of 1.7 kcal/mol from coupled-cluster references).[67] This functional difference alone would be expected to raise the barrier relative to Cantu et al.'s PBE-based value, independent of any finite-temperature or sampling effects. Second, Cantu's CI-NEB barrier is a 0 K potential-energy quantity along a single minimum-energy path, whereas the metadynamics-derived barrier reported here is a finite temperature free energy. Cantu et al. describe Reaction 1 as proceeding through a dissociative mechanism, and the sequential pathway found here is dissociative at each step: a water molecule fully departs before the incoming BDC oxygen binds, both for EB1 and for EB2. Third, the two collective variables used here ($Cr\text{-}O_{water}$ and $Cr\text{-}O_{BDC}$ coordination numbers) are a coarse-grained, two-dimensional projection of the full configurational space; if the true lowest-free-energy path involves degrees of freedom not well represented by these CVs, the reconstructed free-energy surface may not track the same path CI-NEB found. Fourth, CI-NEB locates the minimum-energy path between two chosen, pre-optimized endpoint structures, while metadynamics explores the broader accessible configurational landscape without assuming that path in advance, resulting in a methodological difference in what each calculation answers. This same difference plausibly also explains why Cantu et al.'s path connects the reactant directly to the bidentate product in one concerted step, while the metadynamics trajectories here resolve a stepwise route with a stable intermediate. Which factor dominates cannot presently be isolated, but the functional difference and the path-exploration difference are the most likely primary contributors to the barrier-magnitude gap.

Lastly, a comment on comparing our results to experimental synthetic conditions. The 300 K used throughout this work is well below reported MIL-101 synthesis temperatures (approximately 413 K for a water/DMF route[67] and 493 K for a water/HF route[28]). This choice follows Cantu et al.'s own exploratory AIMD, which also used 300 K, and keeps the present comparison controlled: the fine-tuning workflow is validated against a specific published mechanism before being extended to synthesis-relevant conditions. Extending it to those conditions is not simply a matter of raising the thermostat set point. At 413-493 K, the accessible configurational space broadens through increased solvent exchange and greater thermal expansion of the coordination shell. Representing these synthesis conditions faithfully would also mean modeling the DMF- or HF-containing solvent mixtures rather than the water-only system studied here. Both changes would sharply increase the DFT reference-data burden at the ωB97M-V/def2-TZVPP level of theory used to fine-tune MACE-POLAR-M.

# 4 Conclusion

This work analyzes the first step of MIL-101(Cr) secondary building unit formation using MLIPs. Fine-tuning MACE-POLAR-M on a compact DFT reference dataset (ωB97M-V/def2-TZVPP) produced the first converged 2D free-energy surface for this step, at a higher level of theory than the original PBE treatment of this reaction in literature. Fine-tuning was essential to this result: the pre-trained model predicted an endothermic reaction and an unphysical global minimum, while both fine-tuned models carried to production, Naïve and MultiHead, corrected this to the correct exothermic thermodynamics and the expected Cr-octahedral product basin. This agrees with the overall mechanism from literature, though the free-energy surfaces obtained here show a single water molecule released first rather than two (Section 3.3). We find that fine-tuning is most successful when the configurations added are relevant to the FES prediction. Our work shows a strategy to identify appropriate structures for fine-tuning MLIPs, a key step to obtain free-energy surfaces for MOF formation. As a secondary test, none of the pre-trained or fine-tuned models reproduced the DFT-level binding energy of Reaction 1 (SI Section S5.5), consistent with the exclusion of dissociated fragments from the fine-tuning data.

## Data Availability

Data for this article, including source code, instructions, and tutorial material, are available at MIL-101-Paper_4 at https://github.com/omendibleba/MIL-101-Paper_4.

## Acknowledgments

This work was performed using the computational resources provided by the Notre Dame Center for Research Computing (NDCRC). Y.C. acknowledges funding support from the U.S. Department of Education (award no. P200A210048) and the University of Notre Dame. This work used DeltaAI at University of Illinois Urbana-Champaign (UIUC) through allocation CHM250057 from the Advanced Cyberinfrastructure Coordination Ecosystem: Services & Support (ACCESS) program[68], which is supported by National Science Foundation grants #2138259, #2138286, #2138307, #2137603, and #2138296

## Author Declarations

### Conflict of interest

The authors have no conflict to disclose.